\documentclass[twocolumn]{aastex701}
\shorttitle{Coronal Vector Magnetic Field Observations}
\shortauthors{Paraschiv et. al}
\accepted{Sep. 28, 2026}
\submitjournal{The Astrophysical Journal Letters}

\begin{document}

\title{A First Global Full-Vector Coronal Magnetic Field Inference Using MLSO CoMP and UCoMP Observations}

\author[0000-0002-3491-1983]{Alin Razvan Paraschiv}
\affiliation{National Solar Observatory, Boulder CO 80303, USA}
\email{arparaschiv@nso.edu}

\author[0000-0002-4973-0018]{Zihao Yang}
\affiliation{High Altitude Observatory, NSF NCAR, Boulder, CO, 80301, USA}
\email{zihao@ucar.edu}

\author[0000-0003-2123-6605]{Daniela Adriana Lacatus}
\affiliation{High Altitude Observatory, NSF NCAR, Boulder, CO, 80301, USA}
\email{lacatus@ucar.edu}

\author[0000-0003-0583-0516]{Momchil Molnar}
\affiliation{Southwest Research Institute, Boulder, CO, 80301, USA}
\email{momchil.molnar@swri.org}

\begin{abstract}
Building on recent advances in coronal spectropolarimetry, seismology, and magnetic-field inferences of \citet{yang2024} and \citet{parsolo2026}, we present the first application of the CLEDB $IQUD$ inversion methodology to Mauna Loa Solar Observatory CoMP and UCoMP observations. Using spectropolarimetric $IQU$ measurements together with coronal seismology derived magnetic field B$_{\text{POS}}$ constraints, we recover vector magnetic field solutions across the off-limb solar corona accessible to the observations, demonstrating the feasibility of routine large-scale coronal vector magnetometry without the need for Stokes $V$ observations.
\end{abstract}


\section{Introduction}\label{sec:intro}

This letter applies a newly validated method that infers vector (3D) coronal magnetic fields from global Stokes $IQU$-only observations of the highly ionized corona. We present a qualitative interpretation of this first inference, alongside a compact discussion of the caveats, limitations, and future prospects.


The theoretical framework to study the polarization of coronal near-infrared (NIR) emission lines resulting from forbidden atomic transitions has been continuously refined. A thorough review of coronal ions and their magnetic diagnostics is presented by \citet{Casini+2017}. The work of \citet{judge2007} established that the \ion{Fe}{13} coronal 10747~\AA{} and 10798~\AA{} lines are optimal for diagnosing the coronal magnetic fields. Following, multiple coronal magnetic field inversion schemes were proposed, largely built on the single-point approximation, which requires that the line-of-sight (LOS) integration of an optically-thin observation captures a dominant and homogeneous structure. This approximation cannot be considered universal when diagnosing the global corona. Analytical inversion schemes were devised by \citet{Plowman2014} and \citet{Dima+2020}. Alternatively, \citet{Kramar+2016}, proposed tomographic approaches that can account for the effects of complex LOS integrations of the optically thin coronal plasma. 

Analytical inversion methods suffer from degeneracies due to an implicit assumption that the field comes strictly from the plane-of-sky (POS) as extensively discussed by \citet{Dima+2020}, limiting their practical applicability. Relaxing the POS magnetic field assumption requires exploring additional realizations of observation geometries that cannot be analytically parametrized. \citet{Judge+Casini+Paraschiv2021} argued that the degeneracies can be lifted by using atomic alignment to constrain the LOS position of the assumed dominating emission source. The development of the CLEDB inversion \citep{par+judge2022} represents a practical application of this principle. Through CLEDB, the observation geometry is constrained by comparing the observation with a database of theoretical Stokes $IQUV$ calculations in which a grid of observational geometries are discretized. This allows for inverting magnetic fields from a set of full Stokes $IQUV$ observations.

Coronal Stokes $IQU$ provide constraints on the magnetic orientation only, while the hard-to-measure Stokes $V$ adds a LOS constraint on the magnetic field strength. In a major achievement, \citet{schad2024} utilized state of the art DKIST \citep{rimmele2020} Cryo-NIRSP \citep{fehlmann2016} observations to unequivocally demonstrate the coronal diagnostic potential and the viability of measuring Stokes $IQUV$ of the \ion{Fe}{13} lines. Smaller-aperture coronal polarimetric instruments do not reach the sensitivity required for measuring Stokes $V$. Nevertheless, they have enabled an alternative approach to infer the coronal magnetic field through its POS component using the two-dimensional coronal seismology technique. This approach exploits Alfv\'enic wave signatures observed in the Doppler velocity of the \ion{Fe}{13} coronal lines \citep{tomczyk2007}. \citet{Sharma2023}, together with recent studies enabled by DKIST/Cryo-NIRSP observations \citep{morton2025a,morton2025b,molnar2026}, demonstrated that the corona is permeated by nearly incompressible fluctuations, most commonly interpreted as Alfv\'{e}nic waves, finding excess power in the 3-5mHz range. These results provided new insight into their frequency spectrum, energy transport, and possible origins. The improved temporal resolution and sensitivity of such new-generation observations have expanded the accessible parameter space, revealing that coronal wave signatures contain information beyond simple detections of transverse motions. 

By tracking such wave motions, proxies for the POS magnetic field orientation such as the local wave propagation direction, and the wave phase speed can be determined \citep{yang2020+, yang2020+b}. When combined with electron density diagnostics from the \ion{Fe}{13} line-ratio, the POS-projected magnetic field strength can be estimated \citep{morton+2016,yang2020+b}. In this way, wave-based inferences of global maps of B$_{\mathrm{POS}}$, offer a practical and routinely applicable partial magnetic diagnostic of the global corona. Its observational feasibility has now been demonstrated by measuring the global coronal B$_{\mathrm{POS}}$ over an eight-month period \citep{yang2024}. The wave-based B$_{\mathrm{POS}}$ provides important complementary constraints for interpreting the magnetic field orientation inferences from easier-to-measure Stokes $IQU$-only observations.

In Sec.~\ref{sec:method} we present a methodological summary for applying a newly validated inversion proposed by \citet{parsolo2026} to \ion{Fe}{13} coronal observations, along with a compact interpretation of degeneracies. The inferences of vector magnetic fields in the corona are presented and compared to coronal magnetic field extrapolations in Sect.~\ref{sec:results}. A compact discussion on the significance, reliability, and the most important limitations is pursued in Sect.~\ref{sec:discussion}.

\section{Methodological Sketch}\label{sec:method}

\begin{figure}
\includegraphics[width=\linewidth]{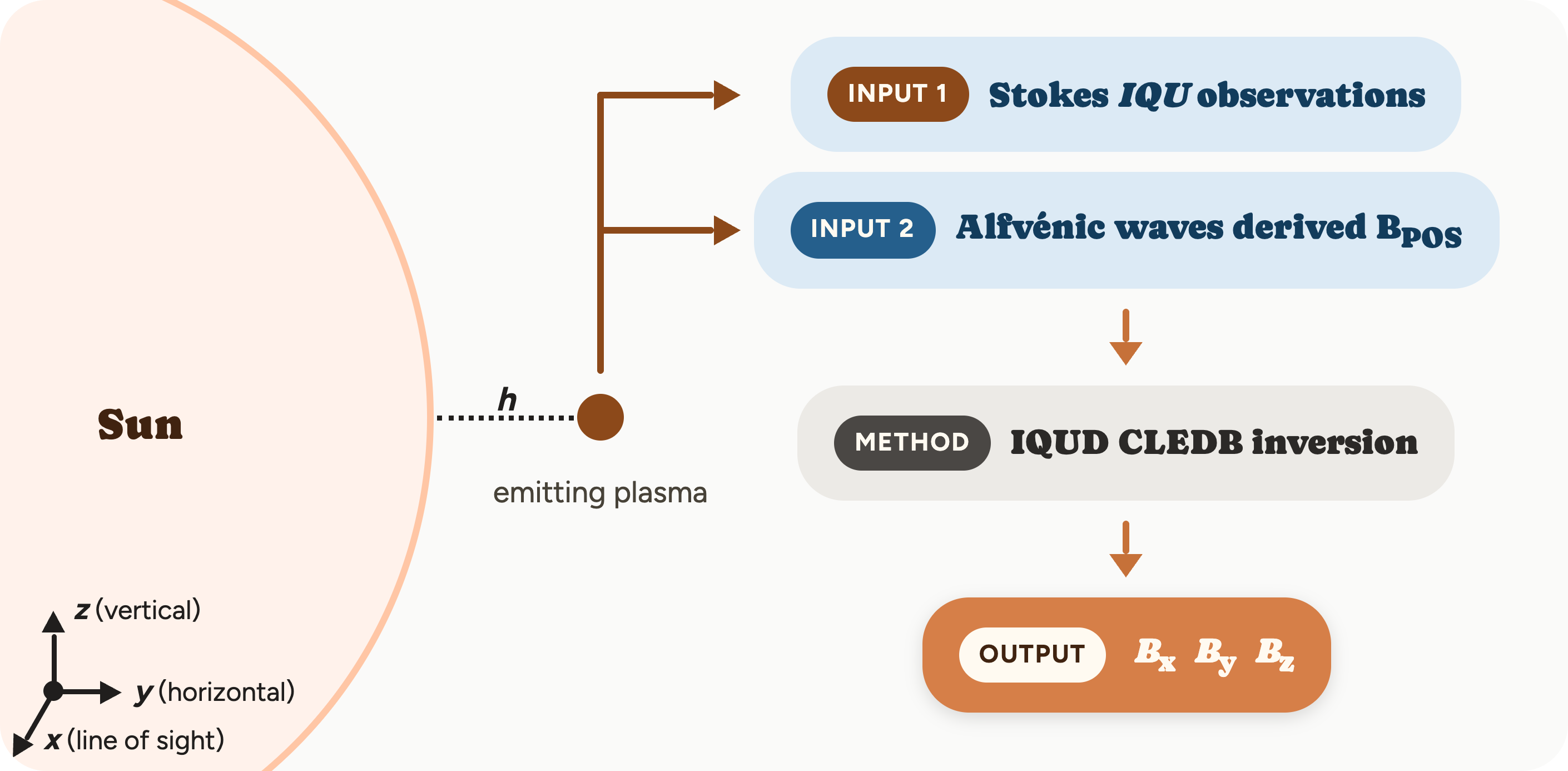}
\caption{To invert vector magnetic fields from an emitting coronal plasma pixel, at an apparent height $h$ above the photosphere, we require two inputs: Stokes $IQU$ observations \citep[see ][]{par+judge2022} and Alfv\'enic waves-derived B$_{\text{POS}}$ \citep[as demonstrated by][]{yang2020+,yang2024}. The inputs are then ingested by the $IQUD$ CLEDB inversion \citep[described and validated by][]{parsolo2026} to infer vector magnetic field components. Our Earth-observing geometric system is defined along $y$ (horizontal) and $z$  (vertical) directions in the plane-of-sky, while $x$ goes along the line-of-sight.}
\label{fig:diagram}
\end{figure}

The \ion{Fe}{13} 10747~\AA{} and 10798~\AA{} spectropolarimetric data used in this work for both waves and CLEDB inversions were acquired by two instruments: the Coronal Multi-channel Polarimeter \citep[CoMP;][]{tomczyk+2008} and an upgraded instrument \citep[UCoMP;][]{2019shin.confE.131T}, deployed at the Mauna Loa Solar Observatory (MLSO). Both perform stable high-cadence coronal observations with a 0.2 m aperture. This approach is particularly promising for the planned 1.5 m aperture COronal Solar Magnetism Observatory \citep[COSMO;][]{cosmo2022,cosmo2023} facility, provisioned to advance our coronal diagnostic capabilities and measure Stokes V over a global field-fo-view (FOV). 

To obtain the magnetic field inferences we utilized the CLEDB inversion method and software package described by \citet{par+judge2022}. This method takes into account the atomic alignment factor ($\sigma^2_0$) to localize the dominant contribution along a LOS integration, to avoid degeneracies resulting from simpler assumptions \citep{Dima+2020,Judge+Casini+Paraschiv2021}. The Alfv\'enic waves based inversion for the plane-of-sky projection of the magnetic field (B$_{\text{POS}}$) have been prototyped and demonstrated by \citet{yang2020+,yang2020+b}. \citet{yang2024} and \citet{yang2026} further demonstrated the improved fidelity of wave angles as tracers of the magnetic field.   

We propose applying the new $IQUD$ inversion algorithm (Stokes $IQU$ + Doppler inferred B$_{\text{POS}}$) implemented in CLEDB, following the theoretical formulation and statistical validation recently presented by \citet{parsolo2026}. We follow the conceptual sketch presented in Fig.~\ref{fig:diagram} to perform inversions of sets of level-2 $IQU$ spectropolarimetric data acquired by the CoMP \citep{comp1074data,comp1079data} and UCoMP\citep{ucompdata} instruments. The $IQUD$ inversions require B$_{\text{POS}}$ estimations as inputs. For this demonstration, we directly utilize previously validated B$_{\text{POS}}$ maps from  2016 Oct. 14, and 2022 Jun. 01, resulting from the analysis described in the \citet[Fig.~3]{yang2020+} and \citet[Fig.~1]{yang2024} studies. Expanded methodological and observational details for the CLEDB inversion, the B$_{\text{POS}}$ wave-tracking approach, and the newly introduced 
$IQUD$ inversion setup, together with the CLEDB software and B$_{\text{POS}}$
datasets used in this work, can be accessed via each respective study.

As described in \citet{parsolo2026}, the current implementation of the CLEDB $IQUD$ inversion will return a set of four degenerate solutions with respect to $\pm\text{B}_x$ and $\pm\text{B}_z$ components.  In this work, the four-fold magnetic field ambiguity was qualitatively resolved for visualization purposes, using an iterative local optimization scheme based on the Iterated Conditional Modes (ICM) algorithm. For each spatial pixel, the inversion solutions were treated as possible states, out of which one was selected through minimizing an $E_k$ function describing the angular disagreement between the candidates and those of neighboring pixels defined as,
\begin{equation}
    E_k = \sum_n^8w_n(1-\hat{\mathrm{b}}_k\cdot\hat{\mathrm{b}}_n),
\end{equation}
where $\hat{\mathbf{b}}_k$ is the normalized candidate vector, $n$ sums over the 8 neighboring $\hat{\mathbf{b}}_n$ normalized vectors, and $w_n$ represents a weighted distance to neighboring pixels. The ambiguity assignments were updated iteratively using an asynchronous (Gauss–Seidel) scheme. The scheme is iterated until solution assignments converged, with no further changes occurring across the map. This approach favors spatially coherent magnetic field orientation clusters while fully preserving the local inversion solutions, but importantly, it does not impose any physical constraints on the CLEDB inverted magnetic field components. We stress that this approach should be considered a \emph{qualitative degenerate solution selection} that cannot replace a thorough physically-based disambiguation approach. 

\section{Results}
\label{sec:results}



\begin{figure*}
\centering
\includegraphics[width=\linewidth]{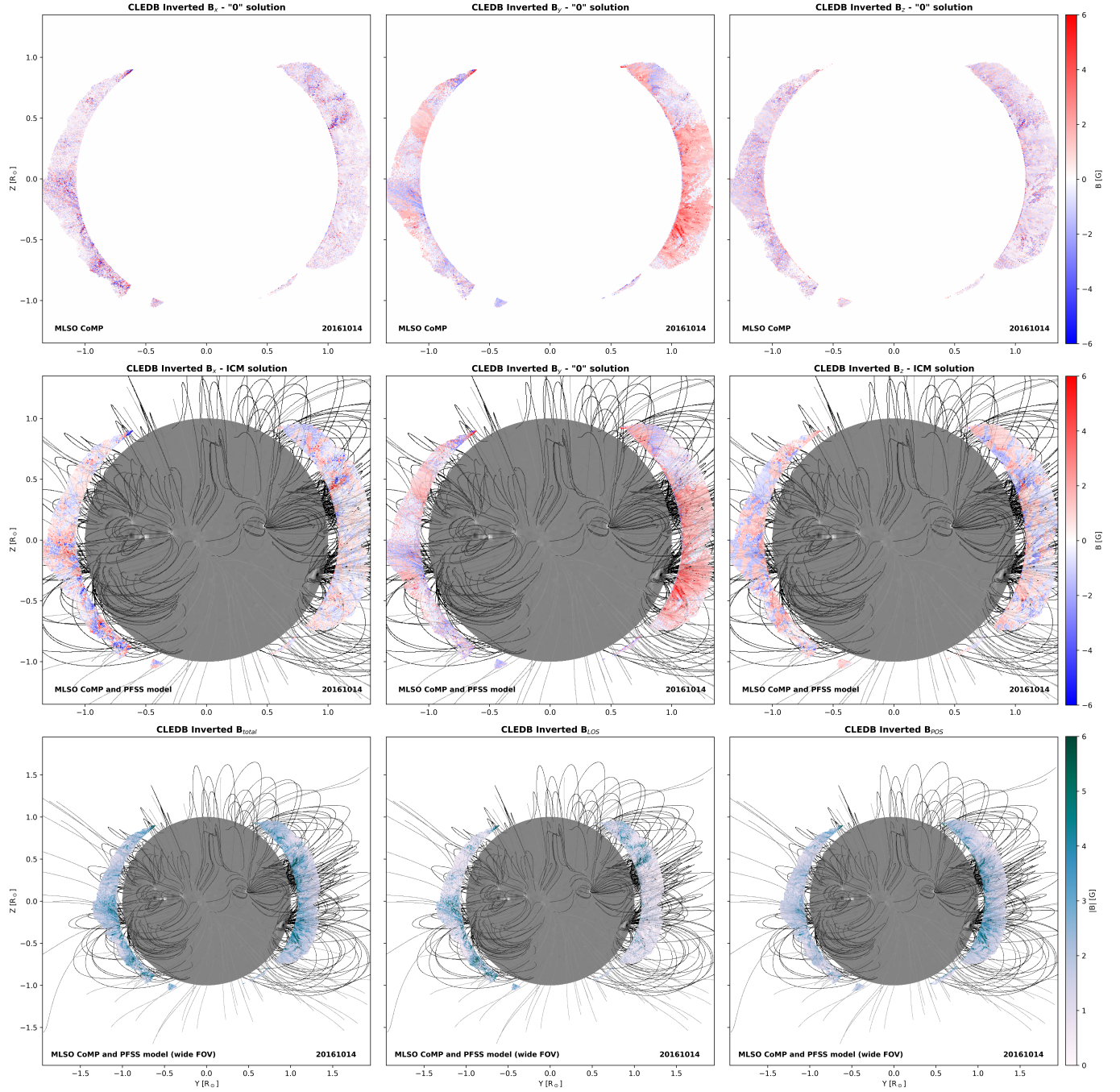}
\caption{CoMP inversions of the vector coronal magnetic field on 2016 Oct. 14. Top row panels present one of the four degenerate solutions. Due to the degeneracies in the sign of B$_x$ and B$_z$ that govern each pixel individually, a noise pattern appears in this distinct solution realization. Qualitative ICM selected realization using the four degenerate solutions are presented in the Middle row. PFSS field-lines derived from photospheric magnetograms are over-plotted. The bottom row presents total magnetic field strength, along with LOS and POS magnitudes in a wider context FOV.}
\label{fig:compinv}
\end{figure*}

Through Fig.~\ref{fig:compinv} we showcase a first mapping of the  inverted coronal magnetic field, using the 2016 Oct. 14 CoMP observation. The top panel presents the B$_x$, B$_y$, and B$_z$ magnetic field components projected in an Earth-observing geometry (see Fig~\ref{fig:diagram}), from the inverted solution candidate "0" as returned by the $IQUD$ CLEDB inversion. Due to the random ordering of the degenerate solutions, the sign-degenerate B$_x$ and B$_z$ components appear in a noise-like blue-red pattern due to the sign flips. We note that the degeneracies do not influence the retrieved field strengths. Small patches are apparent on the East ($[y,z]\sim[-1.1,0.0]$ R$_\odot$) and North-West ($[y,z]\sim[1.1,0.4]$ R$_\odot$) limbs where B$_x$ magnitudes are somewhat larger than B$_y$ or B$_z$. The regions  correspond to Active Regions (ARs), which are near each limb: AR12597 is rotating a second time onto the visible disk, while in the North-West, AR12598 just rotated off the visible solar disk. Additionally, AR12599, situated near the West limb equator has a predominant North-South orientation, confirmed by the magnetograms that show little B$_x$ magnitudes, but strong B$_y$ and B$_z$ magnitudes. 

We additionally compare the inverted magnetic field maps with potential field from source surface extrapolations (PFSS) derived from photospheric magnetograms using the \citet{schrijver2003} method. The B$_y$ and B$_z$ components are mostly representative of global extrapolated field line bundles expanding along larger open structures at the peripheries of the ARs discussed above. More compact arch structures such as the West equatorial region also show clear POS components expanding and then canceling towards heights where loops develop significant inclination. The B$_x$ component is found to be strongest near the most inclined structures of AR12597 and AR12598. We observe that the inversion did not produce valid magnetograms in either North or South polar coronal hole regions, where CoMP signals are generally low. This is expected as these regions are more diffuse, low-density, and prone to LOS-cancellation.

The highest field strengths are recovered around the two limb ARs. To better represent this, the three components are also transformed into total field magnitudes, along with the LOS and POS magnitudes, avoiding the visual complications produced by the solution degeneracies. Analogous to above, the strongest LOS magnitudes appear co-spatial to the East and North-West ARs as confirmed by the inclined PFSS field lines, while the strongest POS magnitudes appear co-spatial with the West limb AR.

\begin{figure*}
\centering
\includegraphics[width=0.95\linewidth]{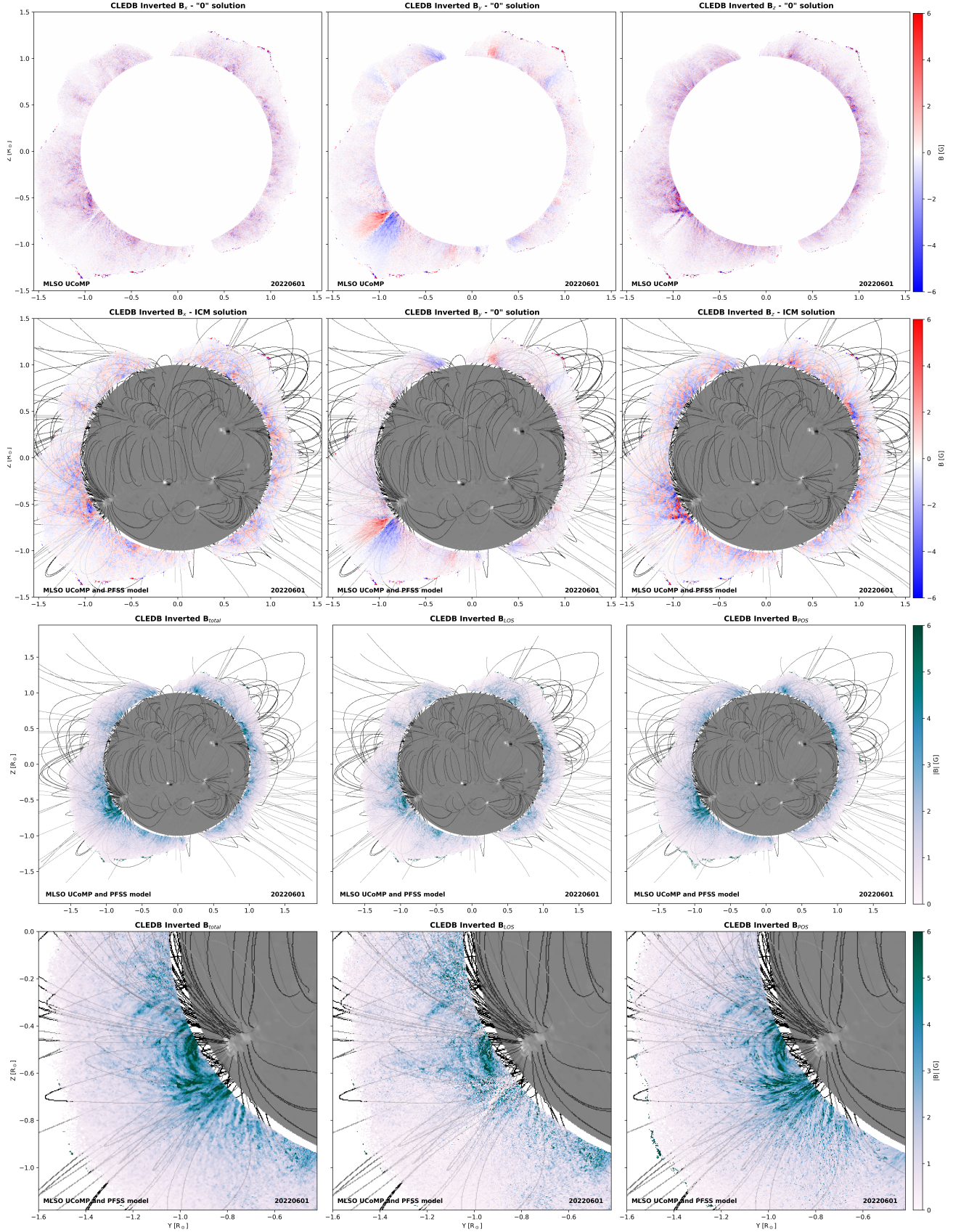}
\caption{UCoMP inversions of the vector coronal magnetic field on 2022 June 01. Top row panels present one of the four degenerate solutions. Due to the degeneracies in the sign of B$_x$ and B$_z$ that govern each pixel individually, a noise pattern appears in this distinct solution realization. Qualitative ICM selected realization using the four degenerate solutions are presented in the second row. PFSS field-lines derived from photospheric magnetograms are over-plotted. The third row presents total magnetic field strength, along with LOS and POS magnitudes in a wider context FOV. The fourth row present zoom-ins of AR13006's total, LOS, and POS magnetic magnitudes.}
\label{fig:ucompinv}
\end{figure*}

The UCoMP maps provide higher-fidelity and higher-quality inferences of the coronal field. This is evident in the 2022 June 1 observation presented in Fig.~\ref{fig:ucompinv}, where individual magnetic structures are more clearly isolated. Comparisons of the LOS and POS magnetic field magnitudes reveals substantial spatial variation in their relative contributions. A prominent example is the strong reversal in B$_y$ around $[y,z]\sim[-1.0,-0.6]$ R$_\odot$, above the decaying AR13006, which has only recently emerged from behind the limb for a second rotation. The AR core appears strongly twisted, with substantial contributions from both the LOS and POS magnitudes. Southward of the core, the fan-like structure is associated with an observed reversal in B$_y$. The corresponding azimuth in Fig.~\ref{fig:azimvsphase} suggests that the field becomes preferentially oriented in a more horizontal direction, with the exception of the strong vertical B$_z$ fields  observed between the B$_y$ reversal polarities. This configuration is confirmed through examination of the Stokes polarization states that reveal a perpendicular direction change across a narrow spatial region. Together, these signatures may indicate the large-scale manifestation of a current sheet oriented approximately along the POS. Consistent with this interpretation, B$_x$ and B$_{\mathrm{LOS}}$ in this region are almost inexistent, whereas a bundle of apparently radial extrapolated field lines expands outward from the same location. The details highlighted in this observation reveal new opportunities for detailed investigations of the three-dimensional magnetic structure of ARs.

We further consider the azimuthal (i.e. along POS) direction of the field given by the combination of B$_y$ and B$_z$. We obtain two independent estimations, one through the linear polarization Stokes $QU$ components and one through the Alfv\'enic wave propagation direction. Fig.~\ref{fig:azimvsphase} presents comparisons for both CoMP and UCoMP. The CoMP maps visually show differences, where finer details are revealed in the wave-angle realization. The differences appear particularly with respect to the $2\pi\rightarrow0$ traversals.  When comparing the UCoMP maps, we find some notable differences. One example is the South-east limb, where the phase speeds show multiple azimuth-reversing structures, while the linear polarization shows broader, more uniform regions. We discuss the origin of these differences in Sec.~\ref{sec:discussion}.

From a theoretical perspective, the two NIR \ion{Fe}{13} emission lines are formed under conditions where photoexcitation contributes significantly alongside collisional excitation. A strong formation dependence with respect to observation height, that is distinct for each line is consequentially manifesting. This effect is relevant when discussing LOS integration, as it creates bias towards structures closer to the POS \citep{parsolo2026,Paraschiv2026b}. This bias is further reinforced by the geometry required to invert the Alfv\'enic waves magnetic field projection, where radial geometries are better constrained in terms of wave propagation \citep{yang2020+b,yang2024}. We will expand upon this aspect in Sec.~\ref{sec:discussion}. In the examples shown in Figs.~\ref{fig:compinv}- \ref{fig:ucompinv}, we find that CoMP produced a reasonable inversion output up to apparent heights of $\sim\;1.15-1.25$ R$_\odot$. The improved sensitivity of UCoMP enabled the recovery of magnetic fields up to greater $\sim\,1.20-1.45$ R$_\odot$ heights, particularly above ARs. The inversion issues trackers implemented in CLEDB show for these two cases that the input Stokes $IQU$ signals above similar respective heights start becoming unreliable due to overall lower signals. More generally, the applicable height ranges recoverable with $IQUD$ CLEDB inversion will vary with the solar cycle, with the coronal structures, and associated LOS complexity contributing towards a particular observation.

\section{Discussion}\label{sec:discussion}

This work presents the first inferences of the vector coronal magnetic fields using CoMP and UCoMP observations using the newly developed and validated CLEDB implementation of a Stokes $IQUD$ inversion that utilizes Alfv\'enic wave constraints. The wave inversion is observationally validated by \citet{yang2020+} and \citet{yang2020+b}. The coronal magnetograms presented in this work have degeneracies embedded in the solutions, analogous to those in photospheric magnetic field inversions. For this scope, we attempted a strictly qualitative solution selection that does not substitute a formal disambiguation. We ground the accuracy of the inversion on the comprehensive analyzes and statistical validations of the methods as described by \citet{yang2020+,par+judge2022,yang2024} and \citet{parsolo2026}. 

When considering the resulting magnetograms, we find that not all pixels yield reliable solutions, with some degree of visual granularity between pixels with higher and lower inversion output quality. This behavior is primarily associated with the spectropolarimetric signal, particularly the \ion{Fe}{13} 10798~\AA{} line, for which the $Q$ and $U$ signals are more difficult to detect above noise levels. In future work, weighting the two-line Stokes $IQU$ components to give greater importance to the 10747~\AA{} line can partially mitigate this limitation, since they provide similar information on the magnetic field orientation. A distinct type of granularity is apparent in the B$_{\mathrm{LOS}}$ and B$_x$ maps. This is because the primary magnetic constraint in the $IQUD$ inversion is provided by the B$_{\mathrm{POS}}$ measurement, while the LOS component is constrained indirectly through the polarization-derived field geometry, atomic alignment, and reconstructed field-strength \citep{parsolo2026}. Nonetheless, we find that the B$_\mathrm{LOS}$ component is substantial in our inferences, considering that the inversion is not constrained by any LOS sensitive Stokes $V$ measurements. Thus, this study represents a first observational confirmation of the importance of accounting for the atomic alignment through LOS discretizations in coronal inversions \citep{Dima+2020,Judge+Casini+Paraschiv2021,par+judge2022}.

The comparisons of the ICM-derived and field magnitude maps with PFSS modeling should be regarded as a qualitative consistency check. While PFSS extrapolations rely on simplifying assumptions that cannot fully capture the complexity of the measured coronal magnetic field, the overall agreement provided independent support for the physical plausibility of the inferred magnetic fields. Devising a quantitative comparison and a more broadly quantifiable coronal disambiguation algorithm applicable to inversion solutions requires extended effort beyond the scope of this case demonstration. We urge dedicated efforts into exploring potential physically constrained functionals for breaking these degeneracies. We propose prioritizing the three potential approaches summarized by \citet{parsolo2026}; i.e. energy minimization, tomography and Stokes $V$ complementarity. The latter is particularly compelling. The CLEDB $IQUV$ and $IQUD$ inversion modes provide complementary constraints on the magnetic field, with the respective constraints acting primarily on either LOS and POS components. Combining these independent constraints can therefore provide a more complete determination of the vector magnetic field and, in favorable cases, allow a fully disambiguated solution to be inferred without external constraints. A joint "$IQUVD$" inversion can also reduce the pixel-to-pixel granularity associated with the weaker LOS constraint found in the results presented here. Disambiguation of the vector coronal field will enable interpretations of state-of-the-art coronal observations and coordinated observing programs designed to study the magnetism of the corona. DKIST Cryo-NIRSP already provides this capability for detailed, small-FOV observations, while the future COSMO observatory is proposed to measure Stokes $V$ over a large coronal FOV. The CORSAIR balloon-borne spectropolarimeter \citep{samra2021corsair,samra2024} similarly targets Stokes $V$ diagnostics as a pathfinder for future space-based instrumentation. FASR \citep{gary2023} radio imaging spectroscopy can provide complementary coronal magnetic diagnostics over large FOVs.

\begin{figure*}
\centering
\includegraphics[width=\linewidth]{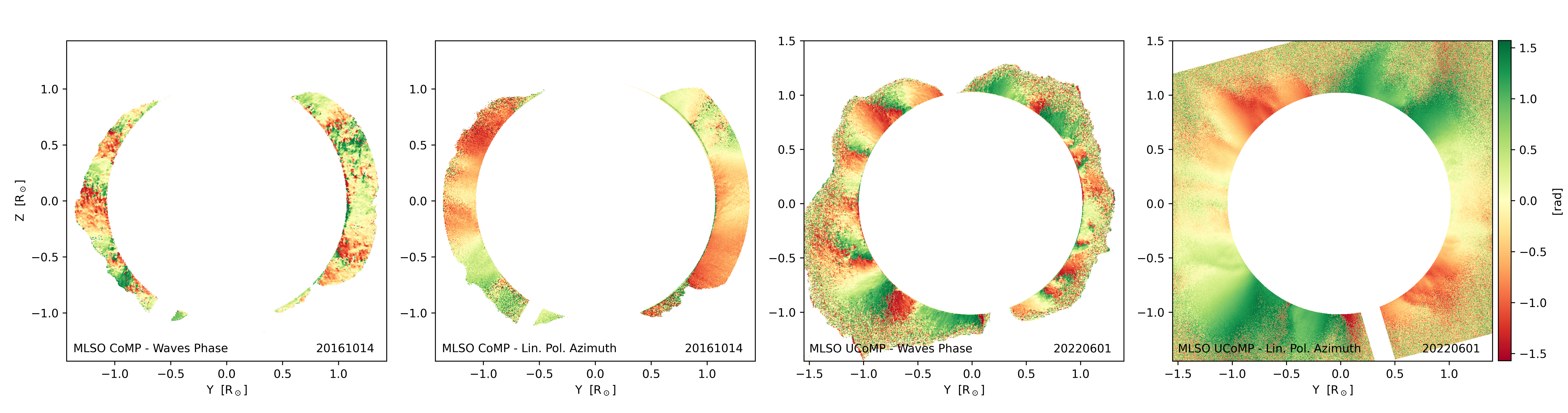}
\caption{The azimuthal orientation of the field is presented through two inferences. The wave phase angles are recovered through the coronal seismology inversion. The Stokes $QU$ polarization azimuth is approximated analytically as $\Phi_B=0.5\arctan(U/Q)$. Both inferences returned comparable magnetic field azimuth orientations for both CoMP and UCoMP observations, considering the distinct degeneracies manifesting in each case.}
\label{fig:azimvsphase}
\end{figure*}

A number of important limitations need to be reiterated. For a statistical confidence, current instruments like CoMP and UCoMP require temporal tracking of Doppler oscillations for time periods of approximately one hour. Thus, $IQUD$ inversions can currently be applied only to static observation use cases. In addition, results including eruptive structures should be subjected to extra skepticism, as the inferred average  density over the one-hour observations needs to be representative of the probed plasma structures. Importantly, the use of magnetic fields diagnosed from coronal seismology is not straightforward. The wave-inferred magnetic field is more accurately described as an LOS emissivity-weighted average rather than a B defined strictly in the POS \citep{yang2024}, corroborating theoretical interpretations \citep{Dima+2020,Judge+Casini+Paraschiv2021,Paraschiv2026b}. This is further complicated in cases where the LOS contributions significantly depart from one coherent and dominating structure assumption (i.e. the single-point). The wave-inferred magnetic field can also be influenced by the distribution of multiple wave amplitude along the LOS, for instance in the polar regions, where the recovered field is canceling along the LOS and becoming weaker than expected. Therefore, while treating the wave-based magnetic field as B$_\mathrm{POS}$ may be appropriate, this assumption will not always hold across the CoMP/UCoMP FOV and could introduce systematic uncertainties. 

The differences observed in the azimuthal maps in Fig.~\ref{fig:azimvsphase} may be explained by several factors: differences in diagnostic ambiguities, the differing temporal requirements of the two methods, the relatively low degree of polarization in one or both lines, and sensitivity to LOS integration. The azimuth inferred from linear polarization is intrinsically affected by the Van Vleck ambiguity, whereas the wave angles derived from wave tracking provide a more accurate estimate of the coronal magnetic field direction and are primarily affected only by the 180$^\circ$ ambiguity \citep{tomczyk2007,yang2020+}. The extended temporal requirements for wave tracking may allow transitory flows or more dynamic structures to manifest, whereas the spectropolarimetric observations provide only temporally-short snapshots. The relatively low polarization state of one or both lines, particularly that of the 10798~\AA{} line might introduce uncertainty, where the relative emissivity factor between the \ion{Fe}{13} lines vary significantly over a relatively shallow range of observation heights, from 1.5 at $\sim1.0R_\odot$ to $10$ at $\sim2.0R_\odot$ \citep{schad+dima2020}. Additionally, the degree of polarization can drop very fast with height, leading to noisy measurements and incorrect Stokes U/Q ratios \citep{schad+dima2020,parsolo2026}, with the weaker signals becoming increasingly susceptible to mixing along the LOS. After accounting for the inherent sign ambiguities, the azimuthal differences remained small in our studied cases, indicating that the polarimetric and wave-based inferences probe similar magnetic orientations. More significant discrepancies may arise in more dynamic coronal environments, potentially leading to partial or full-map incompatibilities between the wave-based and spectropolarimetric inferences, and making the $IQUD$ approach less feasible in such cases. We urge careful consideration of orientation agreement between the two azimuths.

Photospheric magnetic field measurements are becoming increasingly central to operational space-weather forecasting, providing the lower boundary for global models such as WSA-ENLIL \citep{mays2015} and physics-based MHD frameworks such as PSI-MAS \citep{linker1999} and AWSoM \citep{vanderholst2014}. The coronal field directly controls CME evolution, solar-wind structure, magnetic connectivity, and the storage and release of magnetic energy, yet comprehensive observations of its three-dimensional structure are still unavailable \citep{gibson2020}. Recent operational validation has demonstrated that improved time-dependent magnetograms can measurably reduce CME arrival-time errors \citep{ryley2021,mays2025}, while new efforts such as NOAA's SWPC and Space Weather Next together with ESA's Vigil are pursuing improved global and farside photospheric magnetometry. However, these models remain fundamentally dependent on extrapolating the coronal magnetic field from the photospheric boundary. Direct and reliable characterization of the coronal magnetic field, rather than its extrapolation from photospheric boundary data, represents a critical need for space weather forecasting, since it is the coronal field that governs the storage and release of the free magnetic energy responsible for flares and coronal mass ejections \citep{sachdeva2026}. Machine learning and tomographic reconstructions (e.g. SuNeRFs, CROBAR, CoroNeRF, etc.) of the coronal field and structures can greatly benefit from the additional constraints provided by coronal magnetograms \citep[e.g.][etc.]{Kramar+2016,Plowman2021,Jarolim2023,jarolim2026,Hsu2026b}.

Direct coronal magnetic measurements as shown here could therefore provide the missing observational constraints needed to move from boundary-driven extrapolations toward observationally constrained, data-assimilative models of the solar atmosphere and heliosphere. In particular, routine measurements of the coronal vector field, including its strength, topology, and non-potential structure, would provide critical constraints on CME propagation, magnetic connectivity and helicity, and ultimately their space-weather impacts. Considering this vector magnetic field inversion avenue to be open, these results enhance the feasibility of future efforts such as COSMO, FASR, and CORSAIR.

\begin{acknowledgments}
The authors thank the reviewer, editor, and data editor for their thoughtful comments, encouraging assessment, and handling of this work. A.R.P was primarily funded for this work by the National Solar Observatory (NSO), a facility of the National Science Foundation (NSF), operated by the Association of Universities for Research in Astronomy (AURA), Inc., under Cooperative Support Agreement number AST-1400405. This research and scientific software development was partly supported by the NASA Heliophysics Division under grant 80NSSC24K0580. Z.Y. and D.A.L. were supported by the  NSF National Center for Atmospheric Research, which is a major facility sponsored by the NSF under cooperative agreement 1852977. Data is provided courtesy of Mauna Loa Solar Observatory (MLSO), facility operated by the High Altitude Observatory, as part of NCAR, supported by the NSF. The authors would additionally like to acknowledge the fruitful interactions with the MLSO team. A.R.P. utilized Large Language Models \citep{openai2026_gpt57,anthropic2026claude} strictly for assistance with language editing content clarity and coding suggestions.
\end{acknowledgments}

\begin{contribution}
A.R.P. led the effort presented in this work. A.R.P. performed the primary data analysis, ran the spectropolarimetric inversions and drafted the manuscript. Z.Y provided and interpreted the wave inverted B$_{\mathrm{POS}}$ estimations and the magnetic azimuths. D.A.L performed the PFSS extrapolations and comparisons with the coronal magnetograms. M.M contributed to the interpretation of the magnetograms and azimuthal maps. All authors contributed to interpreting the results and improving the manuscript text.
\end{contribution}

%
\facilities{MLSO CoMP, MLSO UCoMP}

\software{CLEDB~V0.7.0 \citep[Code repository: \href{https://github.com/arparaschiv/solar-coronal-inversion/releases/tag/v0.7.0-beta}{github.com/arparaschiv/solar-coronal-inversion}]{par+judge2022,parsolo2026};
Waves inversion analysis codes and datasets \citep[Zenodo]{yangzenodo2020,yangzenodo2024};
Astropy \citep{astropy2022};
Sunpy \citep{sunpy_community2020}; 
Numpy \citep{numpy2020}; 
Scipy \citep{SciPy2020}; 
Matplotlib \citep{matplotlib2007}. 
}

\section*{Data Availability}\label{sec:dataavil}
The CLEDB software is available to the community as an open-source distribution. Releases above CLEDB~V0.7.0 have implementations enabling the $IQUD$ inversion scheme showcased in this application in addition to the more classical $IQUV$ mode. 
We make available a repository of python notebooks, alongside the observation data, B$_\mathrm{POS}$ maps, and extrapolation outputs that are required for running the CoMP and UCoMP inversions presented in this work. These are sufficient to reproduce the results presented. We also include a ``frozen'' (static) build of CLEDB V0.7.0.
The repository is hosted through the Harvard Dataverse at \dataset[doi:10.7910/DVN/Y6M7HJ]{ https://dataverse.harvard.edu/dataset.xhtml?persistentId=doi:10.7910/DVN/Y6M7HJ}. An additional 2012 Mar. 27 CoMP analysis example similar to the 2016 Oct. 2016 results presented here is available as a CLEDB test example.
The wave tracking analysis codes used for generating the B$_\mathrm{POS}$ maps used as inputs for the data analysis presented here can be accessed at \citet[Zeondo]{yangzenodo2020} and \citet[Zenodo]{yangzenodo2024}. A local installation of CLEDB ~V0.7.0+ and a built or downloaded precompiled database are required for running the Harvard Dataverse replication notebook.







\bibliography{bibliography}{}
\bibliographystyle{aasjournalv7}


\end{document}